\documentclass[runningheads]{svmult}

\usepackage{makeidx}   
\usepackage{graphicx}  
\usepackage{subeqnar}  
\usepackage{multicol}  
\usepackage{physprbb}  
\makeindex             

\def\Vec#1{\mbox{\boldmath $#1$}}
\begin{document}
\title*{Note on a Micropolar Gas-Kinetic Theory}
\toctitle{Note on a Micropolar Gas-Kinetic Theory
}
%
%
\titlerunning{Micropolar Gas-Kinetic Theory}
%
\author{Hisao Hayakawa
}
\authorrunning{Hisao Hayakawa}
%
%
\institute{Graduate School of Human and Environmental Studies,
Kyoto University, Kyoto 606-8501, Japan
     }

\maketitle              

\begin{abstract}
The micropolar fluid mechanics and its transport coefficients are derived
from the linearized Boltzmann equation of rotating particles. In the
dilute limit, as expected, transport coefficients relating to
microrotation are not important, but  the results are
useful for the description of collisional granular flow on an inclined
slope.
(This paper will be published in Traffic and Granular Flows 2001 edited
 by Y. Sugiyama and D. E. Wolf(Springer)). 
\end{abstract}

\section{Introduction}

  Micropolar fluids are fluids with micro-structures.  They belong to a class
of fluid with a non-symmetric stress tensor. Micropolar fluids
 consist of rigid, randomly oriented (or spherical) particles 
which have own spins and microrotations
suspended in a viscous medium. The concept of microrotation, 
originally, has been proposed by  Cosserat brothers in the theory of
elasticity\cite{cosserat}.
 Condiff and Dahler\cite{condiff}, and
Eringen\cite{eringen} applied its concept to describe fluids with 
micro-structures in the middle of 60s. 
Recently, a comprehensive textbook on micropolar fluids has been
published\cite{lukaszewicz}. 

Physical examples of micropolar fluids can be seen
in 
many fields
where all of them contain intrinsic polarities.
However, we believe that the most interesting application 
of micropolar fluid dynamics which would have potential is 
to characterize granular flows\cite{rev,corsica,powder01}.
In fact, the granular flow is one of flows which have micro-structure and
rotation of particles. Therefore, it is natural to introduce a fluid model
which contains an equation for microrotation  in addition to 
an equation of velocity. Along this idea,
Kanatani\cite{kanatani} formulated micropolar fluid mechanics  for
granular flows, which contains the angle of repose within the frame
work.
 Kano {\it et al.}\cite{kano} have confirmed the quantitative
validity of a micropolar fluid model in a chute flow of granular particles
from the comparison of their simulation of micropolar fluids 
 with  their
experience. It is worthwhile to introduce
 that the velocity profile obtained from 
the micropolar fluid model\cite{kano} is far from the parabolic curve 
expected  from the conventional Novier-Stokes flow.

Although we do not know how the micropolar fluid model is relevant in
other situations of granular materials, it is worthwhile to
investigate fundamental properties of micropolar fluids from the view
of granular physics (see e.g. \cite{hayakawa2000}). 
Thus,
the micropolar fluid mechanics
 is not a nonsense generalization of the Navier-Stokes model, but
is a physically relevant model which has
 many applications in physical systems. 

Recently, Mitarai {\it et al.}\cite{mitarai} have analyzed the collisonal 
flow of particles on an inclined slope. They  suggested that 
the kinetic theory of  granular particles  is useful to derive micropolar
fluid mechanics, and the fluid equation is quantitatively relevant to 
characterize the granular flow. Their success of the application of
micropolar fluid mechanics to granular flows makes us understand
microscopic origin of the micropolar fluid mechanics. Such attempts
have been discussed within kinetic theory of polyatomic fluids
\cite{mccoy,dahler} and there 
are some applications to granular fluid\cite{lun}, 
but most of their works are not well accepted in these days.

The purpose of this paper is to summarize our current understanding of
kinetic theory of micropolar gases. For the sake of simplicity we restrict
our interest to two-dimensional perfectly rough particles in the dilute limit.
To avoid complication which may be one of reasons for not appealing of
the old kinetic theory of rough particles, we adopt the method of eigenvalue 
analysis of linearized Boltzmann equation.\cite{resibois}

The organization of this paper is as follows. In section 2 we explain
the outline of micropolar fluid mechanics and its eigenvalue analysis.
In section 3, we shortly summarize the classical mechanics of binary
collisions of identical rough disks. In section 4, we explain the
framework of kinetic theory of dilute gases of rough disks. We also
summarize the relation between the eigenvalues of linearized Boltzmann
equation and the transport coefficients. In section 5, we demonstrate how 
to obtain transport coefficients in micropolar fluids. In section 6
we will discuss our results.

\section{Outline of Micropolar Fluid Mechanics and its Eigenvalue Analysis}

\subsection{Outline of Micropolar Fluid Mechanics}

In this section, we explain the outline of micropolar fluid mechanics.
The conventional 
fluid mechanics consists of equations of continuity of mass, linear 
momentum and energy. In the fluid without microstructure the conservation 
of angular momentum is automatically satisfied but it becomes a nontrivial
conservation law in the fluid with microstructure. Therefore we need an extra
equation of angular momentum in micropolar fluid mechanics. 
In collisional flows we can adopt the differential expansion of 
the strain field in the stress
tensor,
 the couple stress and the heat flux. The constitutive equation
 is similar to that in the Newtonian fluid. In this paper we restrict  
our interest to two dimensional flows.

The equation of mass conservation is the same as that in usual fluid
\begin{equation}\label{2.1}
\partial_t \rho+\nabla\cdot(\rho {\bf u})=0
\end{equation}
where $\partial_t=\partial/\partial t$, and
$\rho$ and ${\bf u}=(u_x,u_y,0)$
 are the mass density and the macroscopic velocity 
field, respectively. 

The continuity equation of linear momentum, on the other hand, becomes
\begin{equation}\label{2.2}
\rho D_t{\bf u}=\nabla\cdot\sigma
\end{equation}
where $D_t=\partial_t+({\bf u}\cdot\nabla)$ is Lagrange's derivative. 
Here the stress tensor $\sigma$ contains the asymmetric part. In collisional
flows in two dimension, 
the stress tensor $\sigma_{ij}$ can be expanded by strains
\begin{eqnarray}\label{2.3}
\sigma_{ij}&=&
(-p+\zeta\partial_ku_k)\delta_{ij}
+\mu(\partial_iu_j+\partial_ju_i-\partial_k u_k\delta_{ij}) \nonumber \\
& &+\mu_r(\partial_i u_j-\partial_j u_i-2\epsilon_{kij}w_k),
\end{eqnarray} 
where $\epsilon_{kij}$ and $w_k=w\delta_{k,3}$
 are respectively Eddigton's epsilon and
the microrotation (macroscopic spin field). 
Thus, the micropolar fluid mechanics has the spin viscosity $\mu_r$ except
for the usual viscosity $\mu$ and the bulk viscosity $\zeta$. 

The continuity equation of angular velocity is 
\begin{equation}\label{2.6}
n I D_tw=\partial_jC_{j3}+\epsilon_{3ij}\sigma_{ij},
\end{equation}
where the component of the couple stress $C_{ij}$ 
for two-dimensional micropolar flow is given by
 $C_{13}=\mu_B \partial_xw$, $C_{23}=\mu_B \partial_yw$ and
$C_{33}=0$. $n$ and $I$ are respectively the number density and
the momentum of inertia of each disk. There is the relation between $n$ 
and $\rho$ as $\rho=n m$ with the mass of particle $m$.

The energy conservation law, in general, can be written as
\begin{equation}\label{2.7}
\rho D_te =-\partial_i q_i+\sigma_{ij}\partial_iu_j+C_{ij}\partial_iw_j
-\epsilon_{kij}\sigma_{ij}w_k-\Gamma
\end{equation}
where $e$, $\Gamma$ and $q_i$ are respectively the energy density, 
the dissipation rate by inelastic collisions and the heat flux. Now we may
assume Fourier's law $q_i=-\kappa \partial_i T$ with the heat conductivity
$\kappa$ and the granular temperature. In general, the energy density of
micropolar fluid  
consists of two parts, the translational energy and the rotational energy. 
However, the mechanism of the energy transfer between two parts is 
not simple\cite{luding}.
Here, we are only interested in the total energy density of two parts. 
We also assume $\Gamma=0$ to characterize of behavior of small dissipations. 
We will discuss the complete treatment including the energy transfer between
two parts and the energy loss of inelastic collisions elsewhere.

\subsection{Linearized hydrodynamics and its eigenvalue analysis}

Here let us discuss the behavior of linearized hydrodynamics around
the basic state $\bar \rho$, $\bar e$ and $\bar {\bf U}=0$:
\begin{equation}\label{2.8}
\rho=\bar \rho+\delta \rho, \quad {\bf U}=\delta {\bf U},
\quad e=\bar e+\delta e,
\end{equation}
where ${\bf U}=({\bf u}, w)=(u_x,u_y,w)$
 is the three dimensional velocity-microrotation
vector. Assuming the relation between $\delta e$ and $\delta T$ as 
$\delta e=C_V \delta T$ with the specific heat with constant volume $C_V$,
the set of equations for linearized  hydrodynamics
in Fourier space
  thus  becomes
\begin{eqnarray}\label{2.10}
\partial_t\rho_q&=& -i \bar\rho {\bf q}\cdot {\bf u}_q \nonumber \\
\partial_t {\bf u}_q&=&-i \alpha {\bf q}\cdot \rho_q-i\beta {\bf q}\cdot T_q
-\nu q^2{\bf u}_q 
 -\delta {\bf q}({\bf q}\cdot {\bf u}_q)+
2i \nu_r {\bf q}\times {\bf w}_q \nonumber \\
\partial_t w_q&=& -\hat \mu_B q^2w_q+2\hat\mu_r(i {\bf q}\times {\bf u}_q
-2w_q) \nonumber \\
\partial_t T_q &=& -i \gamma {\bf q}\cdot {\bf u}_q -\eta q^2 T_q 
\end{eqnarray}
where ${\bf w}_q=(0,0,w_q)$
 is the microrotation vector and the suffix $q$ represents the Fourier component of the variable and
\begin{eqnarray}\label{2.11}
\alpha&=&\frac{1}{\rho}\left(\frac{\partial p}{\partial \rho}\right)_T
, \quad \beta = \frac{1}{\rho} 
\left(\frac{\partial p}{\partial T}\right)_{\rho}, \quad
\nu= \frac{\mu+\mu_r}{\rho} \nonumber \\
\delta &=& \frac{1}{\rho}(\zeta-\mu_r), \quad 
\gamma=\frac{T}{\rho C_V}\left(\frac{\partial p}{\partial T}\right)_{\rho},
\quad \eta= \frac{\kappa}{\rho C_V} \nonumber \\
\hat \mu_B&=& \frac{\mu_B}{n I}, \quad \hat \mu_r=\frac{\mu_r}{n I},
\quad \nu_r=\frac{\mu_r}{\rho}.
\end{eqnarray}

When we assume ${\bf q}=q\hat x=(q,0,0)$,
the set of equations (\ref{2.10}) can be summarized as
\begin{equation}\label{2.12}
\partial_t\Vec{\Psi}_q=M_q\cdot \Vec{\Psi}_q,
\end{equation}
where $\Vec{\Psi}_q={}^T(\rho_q,u_{x,q},u_{y,q},w_q,T_q)$ and 
\begin{eqnarray}\label{2.13}
M_q &=& \pmatrix{0 & -i q \bar \rho &0 & 0 & 0 \cr
-i \alpha q & -(\nu+\delta) q^2 & 0 &0 & -i \beta q \cr
0 & 0& -\nu q^2 & -2i \nu_r q & 0 \cr
0 & 0 & -2 i \hat\mu_r q & -\hat\mu_B q^2-4\hat\mu_r & 0 \cr
0 & -i\gamma q & 0 & 0 & -\eta q^2 \cr } . 
\end{eqnarray}
Therefore, the eigen equation can be written as
\begin{equation}\label{2.14}
M_q \Vec{\varphi}_{\alpha}^q=\lambda_{\alpha}^q\Vec\varphi_{\alpha}^q.
\end{equation}
With the aid of five eigenvectors $\varphi_{\alpha}^q$, the solution of 
linearized hydrodynamics (\ref{2.12}) is represented by
\begin{equation}\label{2.16}
\Vec{\Psi}_q(t)=\sum_{\alpha=1}^5 c_{\alpha}^q(t)\Vec{\varphi}_{\alpha}^q,
\end{equation} 
where $c_{\alpha}^q(t)$ behaves as 
$c_{\alpha}^q(t)=c_{\alpha}^q(0)e^{\lambda_{\alpha}^q t}$. 

The eigenvalues of eq.(\ref{2.12}) is connected with 
the transport coefficients, where the eigenvalues are the solution of
\begin{equation}\label{2.17}
det\{M_q-\lambda_{\alpha}^q I\}=0. 
\end{equation}
Although the exact solution of (\ref{2.17}) is difficult to be obtained,
the hydrodynamic solution near $q=0$ can be obtained easily.
Following the method introduced in standard
textbooks (see e.g. \cite{resibois})
, the five eigenvalues are obtained as follows:
The first two eigenvalues are
\begin{equation}\label{2.18}
\lambda_{1,2}^q=\pm i c_s q-\Gamma_s q^2,
\end{equation}
where $c_s=\sqrt{(C_p/C_V)(\partial p/\partial \rho)_{T}}$ with 
the specific heat with constant pressure $C_p$. $\Gamma_s$ is the rate of
sound absorption, which is given by
$\Gamma_s=\frac{1}{\bar\rho}\left(
\zeta+\mu+\kappa(\frac{1}{C_V}-\frac{1}{C_p})\right).
$
The third mode is
\begin{equation}\label{2.20}
\lambda_3^q=-\frac{\mu}{\bar\rho} q^2,
\end{equation}
and the fifth mode is
\begin{equation}\label{2.21}
\lambda_5^q=-\frac{\kappa}{\bar\rho C_p} q^2.
\end{equation}
The fourth mode is the nontrivial mode which represents the relaxation of
the microrotation as
\begin{equation}
\lambda_4^q=-\frac{4\mu_r}{n I}-\frac{1}{n I}(\mu_B+\frac{\mu_r I}{m})q^2.
\end{equation}
We should note that $\lambda_4^q$ contains a constant which is independent of
$q$. This means that the microrotation field cannot be connected with 
the zero eigenvectors of the linearized Boltzmann (or Enskog) equation.

\section{Binary Collisions of Identical Rough Disks}

In this section, we briefly summarize the result of binary collisions of
two-dimensional, identical, circular disks obtained by 
Jenkins and Richman\cite{jenkins85}.

Here we assume that all variables are non-dimensionalized by the diameter 
of the disk $d$, the thermal velocity $c_T\equiv\sqrt{2T_0/m}$, 
where $T_0$ and $m$ are granular temperature and the mass of a particle, 
respectively. The angular velocity
is assumed to be non-dimensionalized by $\omega_T\equiv\sqrt{2T_0/I}$
with $I=md^2/8$.
All dimensionless quantities are specified by hat. Thus, we introduce
\begin{equation}\label{3.0}
\hat x=\frac{x}{d}, \quad \hat{\bf c}=\frac{{\bf c}}{c_T},
\quad \hat\omega={\frac{\omega}{\omega_T}}.
\end{equation}
 which  means that $d \omega$ 
 is reduced to $2\sqrt{2}\hat\omega c_T$.

Let us consider a binary collision in which the translational 
velocity $\hat{\bf c}_1$,
$\hat{\bf c}_2$ and the spin $\hat\omega_1$, 
$\hat\omega_2$  priori to a collision  to become
$\hat{\bf c}_1'$, $\hat{\bf c}_2'$ and $\hat\omega_1'$, $\hat\omega_2'$. 
In order to characterize binary collisions of two identical disks
we introduce two coefficients of restitution as
\begin{equation}\label{3.1}
\hat k\cdot \hat{\bf v}'=-e \hat k \cdot \hat{\bf v}
\quad {\rm and } \quad 
\hat k \times \hat{\bf v}'=-\beta_0 (\hat k\times \hat{\bf v})
\end{equation}
where $\hat k$ is the unit vector connecting the center of the particle 1
with that of 2, and $\hat{\bf v}$
 is the relative velocity of the points of contact defined by
\begin{equation}\label{3.3}
\hat{\bf v}=\hat{\bf g}+\sqrt{2}\hat z\times \hat k(\omega_1+\omega_2)
\quad {\rm with}\quad 
\hat{\bf g}=\hat{\bf c}_1-\hat{\bf c}_2.
\end{equation} 
The coefficients of restitution are not constants in actual situation.
In particular, the tangential restitution coefficient $\beta_0$ strongly
depends on the ratio of the tangential velocity to the normal velocity 
when the Coulomb slip takes place.
However,
 they may be regarded as constants in a wide range 
approximately. The coefficient
$e$ may take values from 0 to 1, while $\beta_0$ may vary between -1 and 1.
The elastic hard core collision is characterized by $e=1$ and $\beta_0=-1$ but
the collisions of perfectly rough disks are characterized by $e=\beta_0=1$,
where the energy is conserved in the binary collisions and there is
still the  time reversal
symmetry.

In this paper, we are interested in the case $e=\beta_0=1$, because 
this situation allows the simplified description with time reversal symmetry
but the spin can play nontrivial roles. In this situation, we have
the changes in a binary collision of the relative velocity $\hat{\bf g}$ 
\begin{equation}\label{3.5}
\hat{\bf g}'-\hat{\bf g}=-\frac{2}{3}\hat{\bf v}-\frac{4}{3}\hat k(\hat{\bf g}\cdot\hat k),
\end{equation}
the velocity
\begin{equation}\label{3.6}
\hat{\bf c}_1'-\hat{\bf c}_1= -\frac{\hat{\bf v}}{3}-\frac{2}{3}
\hat k(\hat k\cdot\hat{\bf g}), \qquad
\hat{\bf c}_2'-\hat{\bf c}_2=\frac{\hat{\bf v}}{3}+\frac{2}{3}\hat k(\hat k\cdot{\bf g}),
\end{equation}
and  the spin 
\begin{equation}
\label{3.7}
\hat\omega_1'-\hat\omega_1= 
-\frac{\sqrt{2}}{3}(\hat k\times \hat{\bf v})\hat z, 
\qquad
\hat\omega_2'-\hat\omega_2= -\frac{\sqrt{2}}{3}
(\hat k\times \hat{\bf v})\hat z.
\end{equation}
More general results for any $e$ and $\beta_0$ can be seen in the paper 
by Jenkins and Richman\cite{jenkins85}. For example, let us present one
result for any $\beta_0$ as 
\begin{equation}\label{3.8}
\hat\omega_1^*+\hat\omega_2^*-\hat\omega_1-\hat\omega_2=
-\frac{\sqrt{2}}{3\beta_0}
(1+\beta_0)\hat z\cdot(\hat k\times \hat{\bf v}),
\end{equation}
where $\hat\omega_i^*$ denotes the spin of $i$ priori to a collision to become 
$\hat\omega_i$. Note that there is time reversal symmetry for any $e$ and
$\beta_0$.

\section{Kinetic Theory of Dilute Gases}


In this section, we discuss the kinetic theory of dilute rough
disks. Since the Boltzmann equation should have modification if
there is no time reversal symmetry, here we assume that the disks
are perfectly rough, i.e.  $e=\beta_0=1$ to keep the time reversal symmetry.
As will be shown, the result recovers usual Navier-Stokes equation in
the dilute limit, but in some situation like in flows on an inclined
rough plate the prediction of 
the dilute gas kinetics seems to be quantitatively useful\cite{mitarai}. 
The extension of this analysis for dense particles with dissipative
collisions will be discussed elsewhere.
Note that there are some treatments of three dimensional dissipation-less
polyatomic
fluids\cite{mccoy,dahler} based on Chapman-Enskog scheme.  On the other
hand, Jenkins and Richman\cite{jenkins85} 
developed kinetic theory of rough inelastic
disks in two dimensions based on the Grad expansion. It is difficult to 
check their argument because their calculation is
 long and complicated. In addition, their method
cannot derive micropolar fluid mechanics as a closed form.
Here we introduce a
simpler method of calculation which  can derive micropolar fluid 
mechanics and can determine the transport coefficients.

We begin with Boltzmann equation of particles with diameter $d$:
\begin{equation}\label{4.1}
\partial_tf_1+({\bf c}_1\cdot\nabla)f_1=
d\int d{\bf c}_2\int_{-\infty}^{\infty}d\omega_2
\int d\hat{k} H(\hat{\bf g}\cdot\hat{k}) g 
 \{f'_1f'_2-f_1f_2\}
\end{equation}
Here we consider the collisions $({\bf c}^*,\omega^*)\to ({\bf c},\omega)\to
({\bf c}',\omega')$, but we note $({\bf c}^*,\omega^*)=
({\bf c}',\omega')$ because of time reversal symmetry for $e=\beta_0=1$.
In eq.(\ref{4.1}), we use the following notations: 
$\hat k$ is the common unit normal vector at contact,
$\hat {\bf
g}=\hat{\bf c}_1-\hat{\bf c}_2$, $\hat g=|\hat{\bf g}|$,
$H(x)=1$ for $x\ge 0$ and $H(x)=0$ for $x<0$, 
$f_1'\equiv f({\bf r},{\bf c}'_1,\omega'_1)$ and 
$f_1=f({\bf r},{\bf c}_1,\omega_1)$.

 Now let us analyze a linearly nonequilibrium situation. The distribution
function can be expanded as
\begin{equation}\label{4.2}
f({\bf r},{\bf c},\omega,t)=n({\bf r},t)f_0({\bf c},\omega)
(1+\Psi({\bf r},{\bf c},\omega))   
\end{equation} 
where $n$ is the number density, and  $f_0$ is the Maxwell-Boltzmann 
distribution function which vanishes in the collisional integral in 
eq.(\ref{4.1}).

Now we adopt dimensionless quantities as in the previous section. Here we
only introduce two variables as
\begin{equation}\label{4.3}
\hat n(\hat x)= n(x)d^2, \quad
f_0=\frac{m\sqrt{I}}{(2T_0)^{3/2}}M(\xi)
\end{equation}
where 
$\Vec{\xi}=({\bf c},\hat \omega)$ and $M(\Vec\xi)=
\exp(-\Vec\xi^2)/\pi^{3/2}$. 

 It is easy to show that the zeroth order 
equation is given by $d\hat n/d\hat x=0$.
Thus, the distribution function $f$ depends on the spatial coordinate only 
through $\Psi$. 

The perturbative equation is now reduced to
\begin{equation}\label{4.5}
\partial_t\Psi_1+
(\hat{\bf c}_1\cdot \hat\nabla)\Psi_1=\hat n\int d^3\Vec{\xi}
\int d \hat k H(\hat {\bf g}\cdot \hat k)\hat g
M(\Vec\xi)\{\Psi'_2+\Psi_1'-\Psi_2-\Psi_1\}
\end{equation} 
This equation can be written as
\begin{equation}\label{4.6}
\partial_t\Psi_1+(\hat{\bf c}_1\cdot \hat\nabla)\Psi_1=\hat n L[\Psi_1], 
\end{equation}
where $L[\Psi_1]$ is the collisional integral in eq.(\ref{4.5}).
Introducing Fourier transform we may rewrite (\ref{4.6}) as
\begin{equation}\label{4.7}
\partial_t \Psi_q+i q\hat c_x \Psi_q= \hat n L[ \Psi_q].
\end{equation}
The solution of (\ref{4.7}) can be obtained from the non-Hermitian 
eigenvalue problem
\begin{equation}\label{4.8}
(\hat L-iq \hat c_x) |\Psi_j^q>=\lambda_j^q|\Psi_j^q>.
\end{equation}

Since we are interested in hydrodynamic behavior of eq.(\ref{4.8}),
we adopt the expansion around $q=0$ as
\begin{eqnarray}\label{4.9}
|\Psi_j^q>&=&|\Psi_j^{(0)}>+q |\Psi_j^{(1)}>+q^2|\Psi_j^{(2)}>+\cdots
\nonumber \\
\lambda_j^q&=&\lambda_j^{(0)}+q\lambda_j^{(1)}+q^2\lambda_j^{(2)}+\cdots .
\end{eqnarray}
Substituting (\ref{4.9}) into (\ref{4.7}) we obtain
\begin{equation}\label{4.10}
\hat n L|\Psi_j^{(0)}>=\lambda_j^{(0)}|\Psi_j^{(0)}>.
\end{equation}
Thus, $|\Psi_j^{(0)}>$ is represented by the linear combination 
of five fundamental eigenvectors as:
\begin{equation}\label{4.17}
|\Psi_{\alpha}^{(0)}>=\sum_{\alpha'=1}^5c_{\alpha\alpha'}|\phi_{\alpha'}>,
\end{equation}
where $|\phi_{\alpha}>$ satisfies 
$\hat L|\phi_{\alpha}>=\lambda_{\alpha}^{(0)}|\phi_{\alpha}>$, and 
their explicit expressions are
\begin{eqnarray}\label{4.18}
|\phi_{1}>&=&1,\quad |\phi_{2}>=\hat c_x, \quad |\phi_3>=\hat c_y,
\nonumber \\
 |\phi_4>&=&\hat \omega, \quad 
|\phi_{5}>=\displaystyle\sqrt{\frac{2}{3}}(\Vec{\xi}^2-\frac{2}{3}).
\end{eqnarray}
Since $|\phi_{\alpha}>$ is the degenerated eigenvector, we need
to use $|\Psi_{\alpha}^{(0)}>$.
The determination of $|\Psi_j^{(0)}>$ will be discussed later.
We also assume that the eigenfunctions are orthonormal as
\begin{equation}\label{4.11}
<\Psi_i^{(0)}|\Psi_j^{(0)}>\equiv \int d\Vec{\xi} {\Psi_j}^{(0)}
{\Psi_i}^{(0)}M(\Vec{\xi})=\delta_{ij}
\end{equation}
Therefore we may introduce $\bar \lambda_j^{(0)}$ as
\begin{equation}\label{4.12}
\bar\lambda_j^{(0)}=\hat n<\Psi_j^{(0)}|L|\Psi_j^{(0)}>.
\end{equation}
This $\bar\lambda_j^{(0)}$ is equivalent to $\lambda_j^{(0)}$ if 
$\lambda_j^{(0)}$ is independent of $\Vec{\xi}$. If $\lambda_j^{(0)}$ 
is a function of $\Vec{\xi}$, two eigenvalues are different from each other.
We believe that $\bar\lambda_j^{(0)}$ plays fundamental roles 
in later discussion.

With the aid of (\ref{4.11}) and (\ref{4.12}) we obtain the relations
at the first order:
\begin{equation}\label{4.13}
|\Psi_j^{(1)}>=
\frac{i q \hat c_x+\lambda_j^{(1)}}{\hat n L-\lambda_j^{(0)}}|\Psi_j^{(0)}>
\end{equation}
and
\begin{equation}\label{4.14}
\bar\lambda_j^{(1)}=i<\Psi_j^{(0)}|q \hat c_x|\Psi_j^{(0)}>,
\end{equation}
where we use $L=L^{\dagger}$ and $\hat n<\Psi_4^{(0)}|L=
\lambda_4^{(0)}<\Psi_4^{(0)}|$. 

From (\ref{4.8}), (\ref{4.9})
 and (\ref{4.13}) we obtain the second order correction 
of the eigenvalue as
\begin{equation}\label{4.15}
\lambda_j^{(2)}=-<\Psi_j^{(0)}|(i q \hat c_x+\lambda_j^{(1)})
\frac{1}{\hat n L-\lambda_j^{(0)}}(i q \hat c_x+\lambda_j^{(1)})|\Psi_j^{(0)}>.
\end{equation}

As the eigenvalues $\lambda_{\alpha}^0=0$ are degenerate, we must be
careful in starting from a proper basis that avoids the appearance of
vanishing denominators. As in the case of quantum mechanics, we must
solve exactly eigenvalue problem in the subspace spanned by
$|\phi_{\alpha}>$.
From the comparison of 
(\ref{4.13}) with (\ref{4.17})
we obtain
\begin{equation}\label{4.19}
\sum_{\alpha'=1}^5c_{\alpha\alpha'}[<\phi_{\alpha'}|\hat c_x|\phi_{\alpha''}>-
\tilde\lambda_{\alpha}\delta_{\alpha,\alpha'}]=0,
\end{equation}
where $\tilde\lambda=i\lambda_{\alpha}^{(1)}/q$.
The required condition to exist nontrivial solutions of this equation is
\begin{equation}\label{4.20} 
det[<\phi_{\beta}|c_x|\phi_{\alpha}>-\tilde\lambda\delta_{\alpha\beta}]=0
\end{equation}
Thus, we obtain
\begin{equation}\label{4.21}
\tilde\lambda_1=-\tilde\lambda_2=\frac{1}{2}\displaystyle\sqrt{\frac{5}{3}},
\quad \tilde\lambda_3=\tilde\lambda_4=\tilde\lambda_5=0.
\end{equation}
For later convenience, we introduce $c_0\equiv \tilde\lambda_1$.
Coming back to eq.(\ref{4.19}) we obtain
\begin{eqnarray}\label{4.22}
|\Psi_1^{(0)}>&=& \frac{1}{\sqrt{2}}\left[
\displaystyle\sqrt{\frac{3}{5}}|\phi_1>+|\phi_2>+
\displaystyle\sqrt{\frac{2}{5}}|\phi_5>\right] , \nonumber \\
|\Psi_2^{(0)}>&=& \frac{1}{\sqrt{2}}\left[
\displaystyle\sqrt{\frac{3}{5}}|\phi_1>-|\phi_2>+
\displaystyle\sqrt{\frac{2}{5}}|\phi_5>\right] , \nonumber \\
|\Psi_3^{(0)}>&=& |\phi_3>, \nonumber \\
|\Psi_4^{(0)}>&=& |\phi_4>, \nonumber \\
|\Psi_5^{(0)}>&=& \displaystyle\sqrt{\frac{2}{5}}\left[-|\phi_1>
+\displaystyle\sqrt{\frac{3}{2}}|\phi_5> \right] .
\end{eqnarray}
These $|\Psi_{\alpha}^{(0)}>$ will be used as the basis of perturbative
treatment.

Taking into account (\ref{4.15}) we obtain the eigenvalues up to the
second order as
\begin{eqnarray}\label{4.23}
\lambda_1&=& -i c_0 q+q^2<\Psi_1^{(0)}|(\hat c_x-c_0)\frac{1}{\hat n L}
(\hat c_x-c_0)
|\Psi_1^{(0)}>, \nonumber \\
\lambda_2&=& ic_0 q +q^2<\Psi_2^{(0)}|(\hat c_x+c_0)\frac{1}{\hat n L}
(\hat c_x+c_0)
|\Psi_2^{(0)}>, \nonumber \\
\lambda_3&=& 
q^2<\Psi_3^{(0)}|\hat c_x\frac{1}{\hat n L}\hat c_x|\Psi_3^{(0)}>, \nonumber \\
\lambda_4&=& \lambda_4^{(0)}+
q^2<\Psi_4^{(0)}|\hat c_x\frac{1}{\hat n L}\hat c_x|\Psi_4^{(0)}>, \nonumber \\
\lambda_5&=& q^2<\Psi_5^{(0)}|\hat c_x \frac{1}{\hat n L}\hat c_x|\Psi_5^{(0)}>. 
\end{eqnarray}


Basically, the eigenvalues obtained in this section should be equivalent
to those obtained in section 2 as the linear micropolar hydrodynamics.
We have to note, however, that the results in this section is written in 
dimensionless forms but the results in section 2 include physical
dimensions in terms of the thermal velocity $c_T=\sqrt{2T_0/m}$ and 
the diameter of particles $d$. Thus we obtain the relations
\begin{eqnarray}\label{4.2.1} 
c_s&=&c_0 \displaystyle\sqrt{\frac{2T}{m}}, \nonumber \\
\Gamma_s&=&d \displaystyle\sqrt{\frac{2T_0}{m}}
<\Psi_1^{(0)}|(\hat c_x-c_0)\frac{1}{\hat n L}(\hat c_x-c_0)
|\Psi_1^{(0)},> \nonumber \\
\mu&=& \rho d \displaystyle\sqrt{\frac{2T_0}{m}} 
<\Psi_3^{(0)}|\hat c_x\frac{1}{\hat n L}\hat c_x|\Psi_3^{(0)}> \nonumber \\
\mu_r &=& -\frac{n I}{4d}\displaystyle\sqrt{\frac{2T_0}{m}}\lambda_4^{(0)},
\nonumber \\
\mu_B&=& -\mu_r \frac{I}{m} -d n I \displaystyle\sqrt{\frac{2T_0}{m}}
<\Psi_4^{(0)}|\hat c_x\frac{1}{\hat n L}\hat c_x|\Psi_4^{(0)}> \nonumber \\
\kappa &=& -\rho d C_p
\displaystyle\sqrt{\frac{2T}{m}}
<\Psi_5^{(0)}|\hat c_x\frac{1}{\hat n L}\hat c_x|\Psi_5^{(0)}>.
\end{eqnarray}

\section{Evaluation of transport coefficients}

In this section, we evaluate the transport coefficients $\mu$, $\mu_r$ and
$\mu_B$ among many transport coefficients. Actually we can describe
incompressible micropolar fluid mechanics in terms of these three coefficients.

\subsection{Evaluation of $\mu_r$}
 
As was shown in the previous section, to evaluate $\mu_r$ we have to
obtain $\lambda_4^{(0)}$. We here present the result for any $\beta_0$,
since it is possible to obtain $\lambda_4^{(0)}$ for 
any $\beta_0$ in (\ref{3.1}). 

Substituting (\ref{3.8}) into (\ref{4.5}) we obtain
\begin{equation}\label{5.1.1}
L[\hat\omega]=-\frac{\sqrt{2}(1+\beta_0)}{3\beta_0}\{{\cal{L}}_1[\hat\omega]+
{\cal{L}}_2[\hat\omega]\}=\frac{\lambda_4^{(0)}}{\hat n}\hat\omega ,
\end{equation}
where
\begin{equation}\label{5.1.2}
{\cal{L}}_1[\hat\omega]=-\int d^2\hat{\bf c}_2\int d\hat k 
H(\hat{\bf g}\cdot \hat k)
(\hat{\bf g}\cdot k)M_2(\hat{\bf c})\hat z\cdot(\hat k\times \hat{\bf g})
\end{equation}
and
\begin{equation}\label{5.1.3}
{\cal{L}}_2[\hat\omega]=
\sqrt{2}\hat\omega
\int d^2\hat{\bf c}_2\int d\hat k
H(\hat{\bf g}\cdot \hat k)(\hat{\bf g}\cdot \hat k)M_2(\hat{\bf c})
\end{equation}
with $M_2(\hat{\bf c})=\exp[-\hat c_x^2-\hat c_y^2]/\pi$.
With the aid of formulae for the modified Bessel function $I_0(z)$ 
with zeroth order and the confluent Hypergeometric function $F(a,b;z)$:
\begin{equation}\label{5.1.4}
I_0(z)=\frac{1}{\pi}\int_0^{\pi}dx e^{z\cos x}, \quad
F(3/2,1;c^2)=\frac{4}{\sqrt{\pi}}\int_0^{\infty}dr r^2 I_0(2rc)
\end{equation}
it is possible to show
\begin{equation}\label{5.1.5}
{\cal{L}}_2[\hat\omega]=\frac{\sqrt{2\pi}}{2}F(3/2,1;\hat c^2)e^{-\hat c^2}
\hat \omega,
\quad {\rm and} \quad
{\cal{L}}_1[\hat\omega]=0.
\end{equation}
Therefore, we obtain the eigenvalue
\begin{equation}\label{5.1.7}
\lambda_4^{(0)}=-\frac{\sqrt{\pi}(1+\beta_0)\hat n}{3\beta_0}e^{-\hat c^2}
F(3/2,1;\hat c^2).
\end{equation}
This result states that the eigenvalue depends on the velocity of
particles.

As mentioned in section 4, if the eigenvalue depends on the velocity,
the effective eigenvalue $\bar\lambda_{\alpha}$ defined in (\ref{4.12})
plays important roles. Thus, $\bar \lambda_4^{(0)}$ is evaluated as
\begin{equation}\label{5.1.8}
\bar \lambda_4^{(0)}=
-\frac{\sqrt{\pi}(1+\beta_0)\hat n}{3\beta_0}\int d^3\Vec{\xi}
\frac{e^{-\Vec{\xi}^2}}{\pi^{3/2}}e^{-\hat c^2}F(3/2,1;\hat c^2)\hat \omega^2.
\end{equation}
The result of integration becomes
\begin{equation}\label{5.1.9}
\bar\lambda_4^{(0)}=-\frac{\sqrt{2\pi}(1+\beta_0)}{6\beta_0}\hat n
\end{equation}
where we use $\int_0^{\infty}dc c e^{-2c^2}F(3/2,1;c^2)=1/\sqrt{2}$.
We note that the result (\ref{5.1.9}) is reduced to
\begin{equation}\label{5.1.10}
\bar\lambda_4^{(0)}=-\frac{\sqrt{2\pi}}{3}\hat n
\end{equation}
in the case of $\beta_0=1$.

From (\ref{4.23}), (\ref{4.2.1}) and (\ref{5.1.10}) we obtain
\begin{equation}\label{5.1.11}
\mu_r=\frac{\sqrt{2\pi}n^2}{96}md^3 \displaystyle\sqrt{\frac{2T_0}{m}}
\end{equation}
for $e=\beta_0=1$.
Thus, $\mu_r$ is proportional to $n^2$ as predicted by 
Mitarai et al.\cite{mitarai}.
Therefore, the effect of microrotation is negligible in usual situations
of dilute gases. However, as demonstrated by Mitarai {\it et al.}\cite{mitarai}
the shear near the boundary produces the relevant situation for microrotation.

\subsection{Evaluation of $\mu$}

$\mu$ can be evaluated from the third equation of (\ref{4.2.1}). This
equation may be rewritten as
\begin{equation}\label{5.2.1}
\hat \mu=-\int d\Vec{\xi} \hat c_x\hat c_y \chi^{xy}M(\Vec{\xi})
= -<\chi^{xy}|Y^{xy}> ,
\end{equation}
where $|\chi^{xy}>=\chi^{xy}$ and $|Y^{xy}>=\hat c_x\hat c_y$. $\hat\mu$ is defined 
by
\begin{equation}\label{5.2.2}
\hat\mu=\mu
/(\rho d \displaystyle\sqrt{\frac{2T_0}{m}})
\end{equation}
and $\chi^{xy}$ is the solution of
\begin{equation}\label{5.2.3}
\hat n L[ \chi^{xy}]=\hat c_x\hat c_y  \quad {\rm or}
\quad \hat n L|\chi^{xy}>=|Y^{xy}>.
\end{equation}

Introducing the function $|f^{xy}>$ vertical to $|\phi_{\alpha}>$, it is easy 
to show that $<\bar f|\hat L|\bar f>$ is minimum for 
$|\bar f>=|\chi^{xy}>$
where $|\bar f>=<f^{xy}|Y^{xy}>/<f^{xy}|\hat n L|f^{xy}> |f^{xy}>$. The minimum principle
then becomes
\begin{equation}\label{5.2.5}
<\chi|Y^{xy}>\le <\bar f|\hat n L|\bar f>
=\frac{|<f^{xy}|Y^{xy}>|^2}{<f^{xy}|\hat n L|f^{xy}>}.
\end{equation}
Here we assume the following expansion:
\begin{equation}\label{5.2.6}
|f^{xy}>=\sum_{j=1}^m \beta_j^{(m)}|f_j>,
\end{equation}
where the coefficient $\beta_j^{(m)}$ is determined by the minimum 
condition of the right hand side of (\ref{5.2.5}). These conditions
are summarized as
\begin{equation}\label{5.2.7}
Y_j^{xy}=\sum_{l=1}\beta_l^{(m)}b_{jl}^{xy}, \quad 
<\chi^{xy}|Y^{xy}>=-\frac{1}{\hat n}\sum_{j=1}^m \beta_j^{(m)}Y_j^{xy},
\end{equation}
where $Y_k^{xy}=<f_k|Y^{xy}>$ and $b_{kl}^{xy}=-<f_k|L|f_l>$.

We assume that $|f_j>$ can be represented by
\begin{equation}\label{5.2.7.1}
|f_j>=S_2^{(j-1)}(\hat c^2)\hat c_x\hat c_y,
\end{equation}
where $S_2^{(j-1)}(x)$ is the Sonine polynomial which is defined by
\begin{equation}\label{5.2.7.2}
S_l^{(r)}(x)=\sum_{j=0}^r\frac{(-1)^j\Gamma(l+r+1)}{\Gamma(l+j+1)(r-j)!j!}x^j
\end{equation}
with the Gamma function $\Gamma(x)$. The Sonine polynomials satisfy 
the orthonormality condition
$\int_{0}^{\infty}dxx^le^{-x}S_l^{(r)}(x)S_l^{(r')}(x)=
\frac{\Gamma(r+l+1)}{r!}\delta_{r,r'}.$
In particular, the relations $S_l^{(0)}(x)=1$ is
useful for later discussion. From the orthonomality 
the
condition of $<f^{xy}|\phi_{\alpha}>=
\sum_j^{m}\beta_j^{(m)}<f_j|\phi_{\alpha}>=0$ is automatically satisfied.

Therefore we obtain the expressions for $Y_j^{xy}$ and $b_{kl}^{xy}$ as
\begin{eqnarray}\label{5.2.7.5}
Y_j^{xy}&=&
 \int d\Vec{\xi}\hat c_x\hat c_y S_2^{(j-1)}(\hat c^2)\hat c_x\hat c_y
M(\Vec{\xi}), \nonumber \\
b_{kl}^{xy}&=&
-\int d{\Vec \xi}M(\Vec{\xi})S_2^{(k-1)}(\hat c^2)\hat c_x\hat c_y
L \hat c_x\hat c_y S_2^{(l-1)}(\hat c^2).
\end{eqnarray}
It is easy to calculate $Y_j$ as
\begin{equation}\label{5.2.7.6}
Y_j^{xy}=\frac{1}{4}\delta_{j,1}.
\end{equation}
It is notable that $Y_k^{xy}$ is zero except for $Y_1^{xy}$.

Since it is known that the expansion in terms of Sonnine's polynomial
is fast, the result with  $m=1$ gives a good approximation.
Adopting this approximation (\ref{5.2.7}) becomes
\begin{equation}\label{5.2.8}
<\chi^{xy}|Y^{xy}>=-\frac{\beta_1^{(1)}}{\hat n}Y_1^{xy}; \quad
Y_1^{xy}=\beta_1^{(1)} b_{11}^{xy}.
\end{equation}
Eliminating $\beta_1^{(1)}$ from (\ref{5.2.8}) we obtain
\begin{equation}\label{5.2.9}
<\chi^{xy}|Y^{xy}>=-\frac{{Y_1^{xy}}^2}{\hat n b_{11}^{xy}}.
\end{equation}

From the definition of the operator $L$ $b_{11}^{xy}$ can be written as
\begin{eqnarray}\label{5.2.10}
b_{11}^{xy}&=&\frac{1}{4\pi^3}\int d\Vec{\xi}_1\int d\Vec{\xi}_2
\int_{-\pi/2}^{\pi/2}d\psi\cos\psi \hat g e^{-\Vec{\xi}_1^2-\Vec{\xi}_2^2}
\nonumber \\
& & \times
[\hat c_{1x}'\hat c_{1y}'+\hat c_{2x}'\hat c_{2y}'-
\hat c_{1x}\hat c_{1y}-\hat c_{2x}\hat c_{2y}]^2.
\end{eqnarray}
where $\pi-\psi$ is the angle between $\hat {\bf g}$ and $\hat k$. To
derive (\ref{5.2.10}) we have used that the cross section ($da/d\psi$ 
with the impact parameter $a$)
is $\cos\psi$ and the time reversal symmetry of
collisions. 
  
From (\ref{3.5}) 
and (\ref{5.2.10})
 we get the relation 
\begin{equation}\label{5.2.13}
b_{11}^{xy}=\frac{7}{18}\displaystyle\sqrt{\frac{\pi}{2}}.
\end{equation}
From (\ref{5.2.1}), (\ref{5.2.6}) and (\ref{5.2.9}) we obtain
\begin{equation}
\hat \mu=\frac{9}{56\hat n}\displaystyle\sqrt{\frac{2}{\pi}}.
\end{equation}
From comparison of (\ref{4.2.1}) with the definition of $\hat\mu$, the
final expression of $\mu$ is given by
\begin{equation}
\mu=\frac{9}{28d}\displaystyle\sqrt{\frac{m T_0}{\pi}}.
\end{equation}
Note that this result is deviated from the result in
ref.\cite{jenkins85}.  

\subsection{Evaluation of $\mu_B$}

The method of calculation of $\mu_B$ is similar to that of $\mu$.
Let us introduce 
\begin{equation}\label{5.3.1}
\mu_c=-\int d\Vec{\xi}\hat \omega \hat c_x\chi^{\omega x}M({\Vec \xi})
=-<\chi^{\omega x}|Y^{\omega x}>,
\end{equation}
where $|Y^{\omega x}>=\hat\omega\hat c_x$ and
$|\chi^{\omega x}>=\chi^{\omega x}$ is the solution of
\begin{equation}\label{5.3.2}
\hat n L|\chi^{\omega x}>=\hat |Y^{\omega x}>.
\end{equation}
Similar to the previous section $\hat \mu_c$ is represented by
\begin{equation}\label{5.3.3}
\mu_c=\frac{{Y_1^{\omega x}}^2}{\hat n b_{11}^{\omega x}}
\end{equation}
in the lowest order approximation. Here $Y_{1}^{\omega x}$
and $b_{11}^{\omega x}$ are respectively given by
\begin{equation}
\label{5.3.4}
Y_1^{\omega x}= \int d\Vec{\xi}\hat(\omega \hat c_x)^2 M(\Vec{\xi})
, \quad 
b_{11}^{\omega x}= -\int d\Vec{\xi}M(\Vec{\xi})\hat \omega \hat c_x
L\hat \omega\hat c_x .
\end{equation}
It is easy to show
\begin{equation}\label{5.3.5}
Y_{1}^{\omega x}=\frac{1}{4}.
\end{equation}
On the other hand,  $b_{11}^{\omega x}$ is rewritten as
\begin{eqnarray}\label{5.3.6}
b_{11}^{\omega x}&=&\frac{1}{4\pi^3}\int d\Vec{\xi}_1\int d\Vec{\xi}_2
\int_{-\pi/2}^{\pi/2}d\psi\cos\psi \hat g e^{-\Vec{\xi}_1^2-\Vec{\xi}_2^2}
\nonumber \\
& & \times
[\hat c_{1x}'\hat \omega_{1}'+\hat c_{2x}'\hat \omega_{2}'-
\hat c_{1x}\hat \omega_{1}-\hat c_{2x}\hat \omega_{2}]^2.
\end{eqnarray}
From (\ref{3.6}) and (\ref{3.7}) we obtain 
\begin{equation}\label{5.3.8}
b_{11}^{\omega x}=\frac{11}{12}\displaystyle\sqrt{\frac{\pi}{2}}.
\end{equation}
Thus $\mu_c$ becomes 
\begin{equation}\label{5.3.9}
\mu_c=\frac{3}{44}\displaystyle\sqrt{\frac{2}{\pi}}.
\end{equation}
From (\ref{4.2.1}) and (\ref{5.3.1}) we obtain
\begin{equation}
\label{5.3.10}
\mu_B
=\frac{3}{176\sqrt{\pi}}d\sqrt{m T_0}
-\frac{\sqrt{2\pi}(nd^2)^2}{768}d\sqrt{2m T_0}.
\end{equation}
It is obvious that the second term is negligible in the dilute limit. 

The method of derivation of $\mu_B$ is relatively simple, when we
compare the method based on Chapman-Enskog scheme in which $\mu_B$
becomes the correction term of higher order\cite{mccoy}.
 
\section{Discussion}

Here, we have demonstrated how micropolar fluid mechanics can be
derived from the Boltzmann equation. Of course, the result is reduced to 
Navier-Stokes equation in the dilute limit. In this sense, at least, we
will have to extend our work to the case of Enskog equation if we
believe that the concept of micropolar fluid mechanics is useful. The effect
of dissipation  also plays important roles though we do not discuss it. 

Roughly speaking the effect of microrotation is localized in the boundary 
layer. Therefore it will be important  to analyze simple shear flows. 
Our preliminary result suggests that the micropolar fluid mechanics may
not be enough to discuss the region close to a flat boundary. 
The success by Mitarai {\it et al}.\cite{mitarai}
 may come from
their bumpy boundary condition in which there is no Knudsen's layer in 
the system due to random scattering of particles near the boundary.

There are many problems to be solved. Let us close
the paper with our perspective whether micropolar fluid mechanics is useful.
In the
pessimistic view the analysis presented here is general nonsense, and
micropolar fluid mechanics is useless. On the 
other hand, in the optimistic view, this work will be a milestone to
discuss the fluid motion with microstructure. Although we are not sure which
result we will see in the future, we hope that micropolar fluid
mechanics is a useful concept to characterize the flow of particles.

The author thanks N. Mitarai and H. Nakanishi for fruitful discussion.

%


\begin{thebibliography}{8.}
\addcontentsline{toc}{section}{References}

\bibitem{cosserat} E. and F. Cosserat, The\`orie des Corps D\`eformables
(A. Hermann, Paris, 1909).
\bibitem{condiff} D. W. Condiff and J. S. Dahler, Physics of Fluids
{\bf 7}, 842 (1964).
\bibitem{eringen} A. C. Eringen, J. Math. Mech. {\bf 16}, 1 (1966).
\bibitem{lukaszewicz} G. Lukaszewicz, Micropolar Fluids: Theory and
Applications (Birkh\"auser, Boston, 1999).

\bibitem{rev} H.M. Jaeger, S. R. Nagel and R. B. Behringer,
Rev. Mod. Phys. {\bf 68}, 1259 (1996).
\bibitem{corsica} H.J.Herrmann, J-P. Havi and S. Luding eds.
Physics of Dry Granular Media (Kluwer Academic 1998).

\bibitem{powder01} Y. Kishino eds.
Powders $and$ Grains 2001 (A.A. Balkema Pub., Rotterdam, 2001).


\bibitem{kanatani} K. Kanatani, Trans. Jpn. Soc. Mech. Eng. B {\bf 45},
507, 515 (1979).
\bibitem{kano} J. Kano, A. Shimosaka, and J. Hidaka, J. Soc. Powder Tecnol. 
Jpn. {\bf 33}, 95 (1996).
\bibitem{hayakawa2000} H. Hayakawa, Phys. Rev. E {\bf 61}, 5477 (2000).
\bibitem{mitarai} N. Mitarai, H. Hayakawa and H. Nakanishi, 
to be published in Phys. Rev. Lett. (cond-mat/0108192).
See also their paper in this proceedings.
\bibitem{mccoy} B. J. McCoy, S. I. Sandler and J. S. Dahler, 
J. Chem. Phys. {\bf 45}, 3485 (1966).
\bibitem{dahler} J. S. Dahler and M. Theodosopulu, Adv. Chem. Phys. {\bf 
	31}, 155 (1975).
\bibitem{lun} C. K. K. Lun, J. Fluid Mech. {\bf 233}, 539 (1991).
\bibitem{resibois} P. Resibois and M. de Leener, Classical Kinetic
	Theory of Fluids (John Wiley $\&$ Sons, New York, 1977).
\bibitem{luding} S. Luding, M. Huthmann, S. McNamara and A. Zippelius,
	Phys. Rev. E {\bf 58}, 3416 (1998).
\bibitem{jenkins85} J. T. Jenkins and M. W. Richman, Phys. Fluids, {\bf
	28}, 3485 (1985).



\end{thebibliography}
\end{document}